\RequirePackage{lineno}
\documentclass[twocolumn,showpacs,superscriptaddress,amsmath,amssymb,nofootinbib]{revtex4-2}
\usepackage{graphicx}
\usepackage{subfigure}
\usepackage{subcaption}
\usepackage{epsfig}
\usepackage{overpic}
\usepackage{dcolumn}
\usepackage{ulem}
\usepackage{makecell}
\usepackage{bm}
\usepackage{color}
\usepackage{lineno}
\usepackage{xspace}
\usepackage{multirow}
\usepackage{epstopdf}
\usepackage{xcolor}
\usepackage{soul}
\usepackage{verbatim}
\usepackage{enumitem}
\usepackage{todonotes}
\usepackage{cancel}
\usepackage[toc,page]{appendix}
\usepackage[colorlinks,allcolors=black]{hyperref}
\allowdisplaybreaks[4]
\usepackage{caption}
\usepackage{diagbox}

\newcommand{\iqm}{\affiliation{State Key Laboratory of Nuclear Physics and Technology, Institute of Quantum Matter, South China Normal University, Guangzhou 510006, China}}

\newcommand{\moe}{\affiliation{Key Laboratory of Atomic and Subatomic Structure and Quantum Control (MOE), Guangdong-Hong Kong Joint Laboratory of Quantum Matter, Guangzhou 510006, China}}

\newcommand{\scnt}{\affiliation{Southern Center for Nuclear-Science Theory (SCNT), Institute of Modern Physics, Chinese Academy of Sciences, Huizhou 516000, Guangdong Province, China}}

\newcommand{\gbrc}{\affiliation{Guangdong Basic Research Center of Excellence for Structure and Fundamental Interactions of Matter, Guangdong Provincial Key Laboratory of Nuclear Science, Guangzhou 510006, China}}

\newcommand{\OU}{\affiliation{Research Center for Nuclear Physics (RCNP), Osaka University, Ibaraki 567-0047, Japan}}

\newcommand{\JAEA}{\affiliation{Advanced Science Research Center, Japan Atomic Energy Agency (JAEA), Tokai 319-1195, Japan}}

\newcommand{\Nishogakusha}{\affiliation{Nishogakusha University, 6-16, Sanbancho, Chiyoda, Tokyo 102-8336, Japan}}

\newcommand{\Keio}{\affiliation{Research and Education Center for Natural Sciences, Keio University, Hiyoshi 4-1-1, Yokohama, Kanagawa 223-8521, Japan}}

\newcommand{\SKCMM}{\affiliation{International Institute for Sustainability with Knotted Chiral Meta Matter (SKCM$^2$), Hiroshima University, 1-3-2 Kagamiyama, Higashi-Hiroshima, Hiroshima 739-8511, Japan}}

\begin{document}
\include{def-com}
\title{\boldmath $Z_c(3900)$ in a hadronic molecule and a triangle singularity approach at finite temperature}

\author{Ying Zhang}
\email{yingzhang@m.scnu.edu.cn}
\email{ying@rcnp.osaka-u.ac.jp}
\iqm
\OU
\moe
\gbrc
\author {Atsushi Hosaka}
\email{hosaka@rcnp.osaka-u.ac.jp}
\OU
\JAEA
\author {Qian Wang}
\email{qianwang@m.scnu.edu.cn}
\iqm
\OU
\moe
\scnt
\author {Shigehiro Yasui}
\email{yasuis@keio.jp}
\Nishogakusha
\SKCMM
\Keio
\date{\today}

\begin{abstract}
Studying exotic hadrons is a challenge against the conventional quark model, providing us with a good platform to  deepen our understanding of the strong interaction. An inclusive study of the exotic hadrons in vacuum and at finite temperature is an intriguing approach to shed light on their nature. As a first step, we study the $Z_c(3900)$ 
in both the $D\bar{D}^\ast$ hadronic molecular and the triangle singularity pictures, and discuss the behaviors of $Z_c(3900)$ at finite temperature in these two different pictures. As a result, we show the properties that its mass becomes smaller and its width becomes larger when the temperature increases, which are seen commonly in the hadronic molecular picture and in the triangle singularity picture. The enhanced widths in hot medium indicate that the $Z_c(3900)$ will be dissociated at a sufficiently high temperature. This feature is also reflected by a decrease of the effective couplings in the hadronic molecular picture. It is concluded that, not only in vacuum but also in hot medium, the behaviors of $Z_{c}(3900)$ are similar in the two different interpretations of $Z_c(3900)$.
\end{abstract}
\maketitle
\section{Introduction}
\label{sec:Introduction}
In the past decades, numerous exotic hadrons, which are expected to have a structure beyond one of the expectations of the conventional quark model, have been observed. This provides a good platform to deepen our understanding of quantum chromodynamics (QCD).
Some of them are the missing hadrons predicted by the conventional quark model based on the SU(6) spin-flavor symmetry,
while some of them are multiquark states, 
which cannot be accepted by either a quark-antiquark meson or three quark baryon picture. 
Great efforts have been put forward to understand their nature, for instance, hadronic molecules, compact multiquarks, hybrids, glueballs, and triangle singularities. For reviews, we recommend Refs.~\cite{Chen:2016qju, Hosaka:2016pey, Esposito:2016noz, Ali:2017jda, Olsen:2017bmm, Karliner:2017qhf, Yuan:2018inv, Liu:2019zoy, Brambilla:2019esw, Guo:2019twa, JPAC:2021rxu, Meng:2022ozq}.

Among these exotic hadrons, the first charged charmoniumlike state $Z_c(3900)^\pm$ attracts great attention from the community. 
Because it is observed in the $J/\psi\pi^\pm$ invariant mass distribution~\cite{BESIII:2013ris}, slightly above the $D\bar{D}^\ast$ threshold, and also charged, there is no doubt that the minimal content is four quarks. Its neutral partner $Z_c(3900)^0$ is also found~\cite{BESIII:2015cld}. In the following, the charge property is implicit, and the $Z_c(3900)$ is used for all three of these $Z_c(3900)^\pm$, and $Z_c(3900)^0$ states.
Owing to its near-threshold behavior, the $Z_c(3900)$ is considered as an isospin triplet $D\bar{D}^\ast$ hadronic molecule~\cite{Wang:2013cya, Guo:2013sya, Wilbring:2013cha, He:2013nwa, Dong:2013iqa, Zhang:2013aoa, Aceti:2014uea, Albaladejo:2015lob, Albaladejo:2016jsg, Gong:2016hlt, He:2017lhy, Ortega:2018cnm, Du:2020vwb, Meng:2020ihj, Wang:2020dgr} or a triangle singularity structure~\cite{Wang:2013cya} naturally.  From the theoretical side, the analytic properties of the production amplitude in these two scenarios are significantly  different so that one could distinguish these two pictures. However, 
from the experimental side, only the invariant mass distributions of the observed channels can be achieved so far, which is not sufficient to distinguish these two scenarios~\cite{Pilloni:2016obd}. Even, the inclusion of the triangle diagram for the hadronic molecular picture can also describe the experimental data very well~\cite{Du:2022jjv,Chen:2023def}. 

Also interesting is that all these exotic hadrons are reported firstly in the vacuum case, i.e., either in $e^+e^-$ annihilation or heavy hadron decays. It is attributed to the experimental detector efficiency in the vacuum environment. 
However, in comparison with the $e^+e^-$ collision and $pp$ or $p\bar{p}$ collision, the hadron properties in heavy ion collisions should involve more mechanisms of strong interaction~\cite{Chen:2024eaq,Das:2024vac,ExHIC:2010gcb,ExHIC:2011say}, for instance, the expansion and cooling of the quark-gluon plasma (QGP), the hadronization of the QGP into the hadron resonance gas (HRG), and the expansion of HRG. Recently, the CMS Collaboration made a milestone for the observation of exotic hadrons in heavy ion collision, i.e., they reported the first experimental signal of the exotic candidate $X(3872)$~\cite{CMS:2021znk} in Pb-Pb collision. This indicates that the study of exotic hadrons in heavy ion collisions is feasible. Luckily, the theoretical study of exotic hadrons in heavy ion collisions started earlier than that in experiment, such as the study of the $D_{s0}(2317)$~\cite{Montana:2020lfi,Braaten:2023ciy}, $Z_c(3900)$~\cite{Abreu:2021xpz}, $X(3872)$~\cite{Montana:2022inz,Zhang:2020dwn,Cleven:2019cre} and so on. For reviews, we recommend Refs.~\cite{ExHIC:2011say,Chen:2024eaq}. 

Considering the achievements of  heavy ion experiments, we aim at pinning down the nature of the $Z_c(3900)$ in the nonvacuum environment, especially distinguishing the hadronic molecular scenario and the triangle singularity scenario.  
In the present work, more specifically, we study the line shape of the $Z_c(3900)^\pm$ at various temperatures in the above two pictures. 

The paper is organized as follow: in Sec.~\ref{sec:framework}, we present our framework for the  hadronic molecular and the triangle singularity pictures. The results and discussions follow as Sec.~\ref{sec:results}. 
The summary is at the end as Sec.~\ref{sec:summary}. 
Some details of the calculations are presented in the Appendixes. 

\section{The two scenarios of the $Z_c(3900)$}
\label{sec:framework}
$Z_c(3900)$ was observed in the $J/\psi \pi^{\pm}$ two-body invariant mass distributions in the three-body final states $J/\psi \pi^+ \pi^-$.  
There are various scenarios about its nature: a $D\bar{D}^\ast$ hadronic molecule, a compact tetraquark, the cusp effect, the triangle singularity. For reviews, we recommend Refs.~\cite{Chen:2016qju, Hosaka:2016pey, Esposito:2016noz, Ali:2017jda, Olsen:2017bmm, Karliner:2017qhf, Yuan:2018inv, Liu:2019zoy, Brambilla:2019esw, Guo:2019twa, JPAC:2021rxu, Meng:2022ozq}. Among them, both the triangle singularity scenario and the $D\bar{D}^\ast$
hadronic molecular scenario are related to the two-hadron threshold and have singularity in the amplitude. In this section, we present the framework of these two scenarios both in vacuum and at finite temperature. Owing to the isospin symmetry approximation, two different charged spectrums of $J/\psi \pi^{\pm}$ are regarded to be equal. Therefore, in the following we often ignore the charge index and simply denote $J/\psi \pi$ unless necessary.  

\subsection{The isospin triplet $D\bar{D}^\ast$ hadronic molecular scenario}
\begin{figure}[h]
    \centering
    \includegraphics[width=1\linewidth]{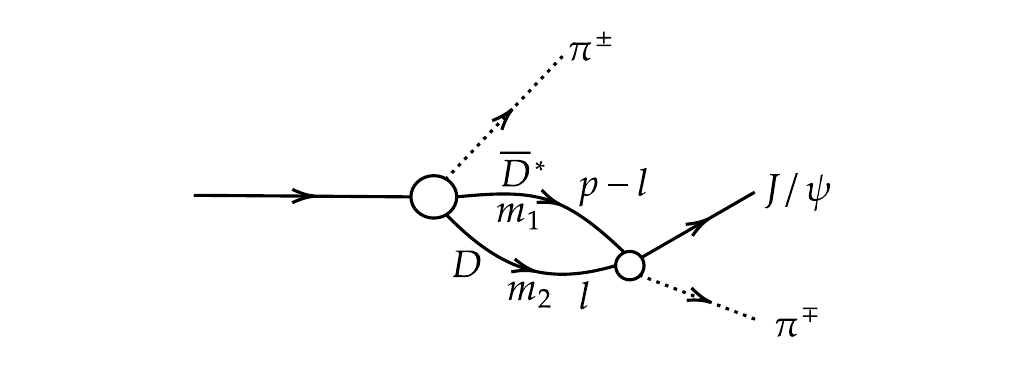}
    \caption{The Feynman diagram corresponding to the production amplitude $\mathcal{M}$ in the hadronic molecular scenario.}
    \label{fig_Zc}
\end{figure}

\begin{figure}[h]
    \centering
    \includegraphics[width=1\linewidth]{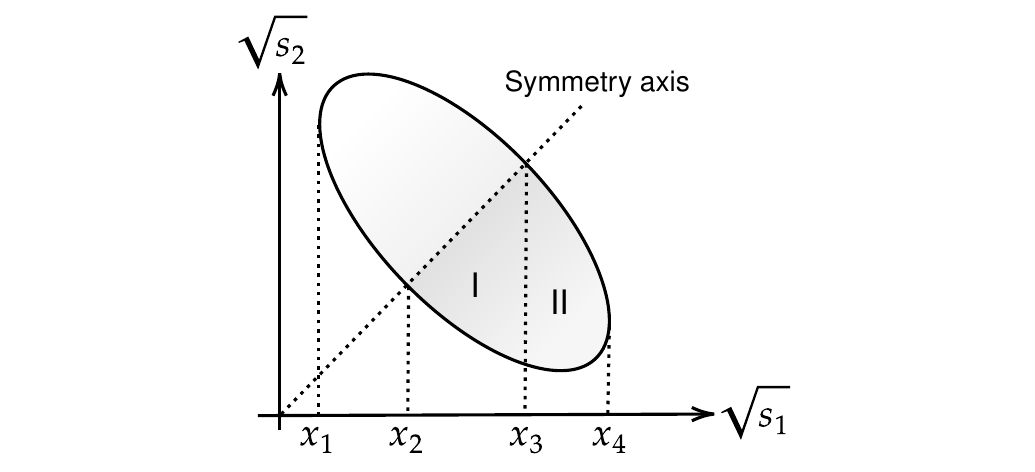}
    \caption{The Dalitz plot (schematic picture).}
    \label{fig_Dalitz}
\end{figure}

The hadronic molecular picture is based on the fact that the Breit-Wigner peak is slightly above the $D\bar{D}^*$ threshold and observed in the $J/\psi\pi$ invariant mass distribution, 
making the $Z_c(3900)$ a good candidate of the isospin triplet $D\bar{D}^*$ hadronic molecule~\cite{Wang:2013cya, Guo:2013sya, Wilbring:2013cha, He:2013nwa, Dong:2013iqa, Zhang:2013aoa, Aceti:2014uea, Albaladejo:2015lob, Albaladejo:2016jsg, Gong:2016hlt, He:2017lhy, Ortega:2018cnm, Du:2020vwb, Meng:2020ihj, Wang:2020dgr}. 
In this case, we consider the $D\bar{D}^*-J/\psi\pi$ two coupled channels. Following the convention of Refs.~\cite{Hanhart:2015cua, Guo:2016bjq}, we use Greek letters $\alpha,\beta\cdots$ for the elastic channel $D\bar{D}^*$ and Latin letters $i,j,k,\cdots$ for the inelastic channel $J/\psi\pi$. 
Owing to the absence of light quarks in $J/\psi$, the direct transition between $J/\psi\pi-J/\psi\pi$ is almost zero~\cite{Hanhart:2015cua, Guo:2016bjq,Liu:2012dv,Detmold:2012pi}, and hence we neglect it here (the same as that in Refs.~\cite{Hanhart:2015cua, Guo:2016bjq}). Thus, the $2\times 2$ Lippmann Schwinger equation can be reduced into two individual equations~\cite{Hanhart:2015cua, Guo:2016bjq},
\begin{eqnarray}
    \begin{cases}
T_{\alpha\beta}=V_{\alpha\beta}+V_{\alpha\gamma}G_\gamma T_{\gamma\beta}+V_{\alpha i}G_iT_{i\beta}\\
T_{i\beta}=V_{i\beta}+V_{i\gamma}G_\gamma T_{\gamma\beta}
    \end{cases},
\end{eqnarray}
and
\begin{eqnarray}
    \begin{cases}
        T_{\alpha i}=V_{\alpha i}+V_{\alpha\gamma}G_\gamma T_{\gamma i}+V_{\alpha j}G_j T_{ji}\\
    T_{ij}=V_{i\gamma}G_\gamma T_{\gamma j}
    \end{cases}.
\end{eqnarray}
Here $G_\alpha, G_\beta,\cdots$ and $G_i,G_j,\cdots$ are
the propagators of the elastic and inelastic channels, respectively. 
By defining the effective potential 
\begin{align}
   V^\mathrm{eff}_{\alpha\beta}\equiv V_{\alpha\beta}+V_{\alpha i}G_i V_{i\beta}, 
\end{align} 
one can obtain the $T$-matrix element 
in the elastic channels
\begin{eqnarray}
T_{\alpha\beta}=V^\mathrm{eff}_{\alpha\beta}+V^\mathrm{eff}_{\alpha\gamma}G_\gamma T_{\gamma\beta}.
 \end{eqnarray}
In our case, only one elastic channel $D\bar{D}^*$ and one inelastic channel $J/\psi\pi$ are of concern, making the above formulas simpler. By parametrizing the transition potential between the elastic channels as a short-range potential denoted as $v$ \footnote{
Here we may ignore the one pion exchange potential (OPEP) between $D$ and $\bar{D}^{\ast}$.  Because of the isospin factor $\vec{\tau}_1 \!\cdot\! \vec{\tau}_2 = 1$  for $Z_c(3900)$ of isospin 1, which is $1/3$ of that for  $X(3872)$ of isospin 0.  The role of the OPEP for $X(3872)$ was discussed, for instance, in Refs.~\cite{Yamaguchi:2019vea,Baru:2015nea,Wang:2017dcq,Voloshin:1976ap, Tornqvist:1991ks, Tornqvist:2004qy, Swanson:2003tb, Braaten:2007ct, Fleming:2007rp, Liu:2008fh}, which is not negligible but not enough to accommodate the $D\bar D^*$ molecule. } and the transition potential between the elastic channel and the inelastic channel as a short-range potential denoted as $g$, the above equations can be simplified as 
\begin{eqnarray}  V^\mathrm{eff}_{11}&=&v+g^2G_2,\\ T_{11}&=&V^\mathrm{eff}_{11}+V^\mathrm{eff}_{11}G_1 T_{11},\\
    T_{21}&=&g+gG_1 T_{11}.
    \label{eqt12}
\end{eqnarray}
Here the labels $1$ and $2$ are used for the elastic $D\bar{D}^*$ and the inelastic $J/\psi\pi$ channels, respectively. For the propagators, we employ the power-divergence subtraction scheme~\cite{Kaplan:1998tg,Kaplan:1998we},
\begin{align}
    &G(E,m_1,m_2)\notag\\
    =&i\int\frac{d^4l}{(2\pi)^4}\frac{1}{(l^2-m_1^2+i\varepsilon)[(p-l)^2-m_2^2+i\varepsilon]}\notag\\
    \simeq&-i\frac{\mu q_0(E,m_1,m_2)}{8\pi m_1m_2},
\end{align}
with the reduced mass $\mu=m_1m_2/(m_1+m_2)$ and the three momentum $q_0(E,m_1,m_2)=\sqrt{2\mu(E-m_1-m_2)}$ in the center-of-mass frame, i.e., $p=(E,\vec{0})$. $E$ is the total energy of the two-body system with masses $m_1$ and $m_2$. The detailed calculations can be found in Appendix~\ref{g2tcal}. The production rate of $J/\psi \pi^+ \pi^-$ can be expressed as a function of the two invariant masses $M_{J/\psi \pi^+}$ and $M_{J/\psi \pi^-}$,
\begin{align}
    d\Gamma
    =&~\frac{1}{(2\pi)^3}\frac{1}{32M^3}|\bar{\mathcal{M}}|^2dM^2_{J/\psi\pi^+}dM^2_{J/\psi\pi^-},
    \label{gamma}
\end{align}
where $M$ is the mass of the initial particle (or the center-of-mass energy $\sqrt{s}$ for the $e^+e^-\to J/\psi\pi^+\pi^-$ scattering process). 
$|\bar{\mathcal{M}}|$ is the module of the spin-averaged amplitude, which contains all the dynamical information.
In the $D \bar D^*$ hadronic molecular scenario, the amplitude goes through 
the $D \bar D^* \pi$ intermediate state where $D \bar D^*$ resonate and 
transit into the final state as $D \bar D^* \to J/\psi \pi$, as shown in Fig.~\ref{fig_Zc}.
For this calculation, we can use the transition amplitude $T_{21}$ in Eq.~(\ref{eqt12}). To obtain the two-body spectrum for $Z_c(3900)$, we integrate over one of the invariant masses in Eq.~(\ref{gamma}), for instance, $M_{J/\psi \pi^-}$, to obtain the spectrum 
as a function of $M_{J/\psi \pi^+}$, i.e., the projection of the Dalitz plot onto 
the $M_{J/\psi \pi^+}$ axis.  
Within our framework, 
since the process $Y(4230)\to J/\psi\pi^+\pi^-$  proceeds via strong interaction, the amplitude $\bar {\cal M}$ is symmetric under the interchange of $M_{J/\psi\pi^+}$ and $M_{J/\psi\pi^-}$, and integrating over $M_{J/\psi\pi^+}$ or $M_{J/\psi\pi^-}$ first will not bring much difference to the result.
In Fig.~\ref{fig_Dalitz}, a schematic Dalitz plot is shown with the notation 
$M_{J/\psi \pi^+} \to \sqrt{s_1}$ and $M_{J/\psi \pi^-} \to \sqrt{s_2}$.
The experimental events of the $1\to 3$ decay process distribute in the whole Dalitz plot as shown in Fig.~\ref{fig_Dalitz}.  
For the $Y(4230)\to J/\psi\pi^+\pi^-$ process, due to the equal masses of the two charged pions, the Dalitz plot is symmetric with respect to the $\sqrt{s_1}=\sqrt{s_2}$ axis. If this process is through strong interaction, that respects isospin symmetry, it will make the dynamics unchanged with the interchange of $\pi^+$ and $\pi^-$. In this case, the event distribution in the whole Dalitz plot is also symmetric with respect to the $\sqrt{s_1}=\sqrt{s_2}$ axis. In this case, the experimental analysis usually projects the events on  the upper left part of the Dalitz region to the lower right part to increase the statistic. For instance, this method is used by the BESIII Collaboration for the analysis of the $Z_c(3900)$ in $J/\psi\pi^+\pi^-$ channel. 
To compare theory predictions and experimental data, we should perform the same procedure in the theoretical side. To project the events on the half Dalitz to $M_{J/\psi\pi^+}$ invariant mass distribution, 
one should perform an integration over  $M_{J/\psi\pi^-}$. 
The lower and upper limits of integration in Region I of Fig.~\ref{fig_Dalitz} are $x_2$ and $x_3$, respectively. Those for Region II of Fig.~\ref{fig_Dalitz} are $x_3$ and $x_4$. 
Here $x_2$ and $x_3$ are the intersection points of the symmetry axis and the boundary of the Dalitz plot. $x_1$ and $x_4$ are the lower limit and upper limit of the Dalitz region, respectively. By plugging the masses of the particles, we find that ($x_2$, $x_3$, $x_4$) $\to$ (3.63, 3.72, 4.12) GeV, respectively. The integration is performed for Region I and Region II separately. The result is
\begin{widetext}
\begin{align}
 \frac{d\Gamma}{dM_{J/\psi\pi^+}}=\begin{cases}
   &\dfrac{1}{(2\pi)^3}\dfrac{1}{32M^3}|\bar{\mathcal{M}}|^22(s_1-s_{2a}+s_{2b})M_{J/\psi\pi^+}~~~(3.63~\text{GeV}<M_{J/\psi\pi^+}<3.72~\text{GeV})\\
   &\dfrac{1}{(2\pi)^3}\dfrac{1}{32M^3}|\bar{\mathcal{M}}|^24s_{2b}M_{J/\psi\pi^+}~~~~~~~~~~~~~~~~~~~~(M_{J/\psi\pi^+}>3.72~\text{GeV})
 \end{cases} ,
 \label{invarHM}
\end{align}
\end{widetext}
where various variables are defined by 
\begin{eqnarray}
 s_1&\equiv& M_{J/\psi\pi^+}^2,\\
 s_{2a}&\equiv& m_{J/\psi}^2+m_{\pi^-}^2\notag\\
 &~&+\frac{(s_1+m_{J/\psi}^2-m_{\pi^+}^2)(s-s_1-m_{\pi^-}^2)}{2s_1},\\
 s_{2b}&\equiv&\frac{\lambda^{1/2}(s_1,m_{J/\psi}^2,m_{\pi^+}^2)\lambda^{1/2}(s,s_1,m_{\pi^-}^2)}{2s_1},
\end{eqnarray}
with $\lambda$ the standard K\"allen function.

To fit the experimental data, the background contribution is added by parametrizing as $\mathrm{b.k.}\equiv\frac{a}{(x-3.6)^b}+c+d x$, the same as that in the experimental work~\cite{BESIII:2013ris}.
Here, $a$, $b$, $c$ and $d$ are the free fit parameters. The fitted formula to the $M_{J/\psi\pi^+}$ invariant mass distribution is
\begin{eqnarray}
    \frac{\mathrm{d}\Gamma}{\mathrm{d}M_{J/\psi\pi^+}}N+\mathrm{b.k.},
    \label{eq:fit_function}
\end{eqnarray}
with $N$ the normalization factor to describe the relative strength between the signal and the background. The inclusion of the background is also used for the triangle singularity scenario as discussed in the next subsection. We use this formula to fit the $M_{J/\psi\pi^+}$ invariant mass distribution in vacuum and make an extension to the invariant mass distribution at finite temperature. 

From the QCD phase diagram, we know for small nucleon density, when the temperature goes above about $150~\mathrm{MeV}$, hadrons are dissociated into quarks and gluons. Our expression in the hadron level does not work anymore. 
Thus, we should work at the temperature below $150~\mathrm{MeV}$. In this temperature region, 
the short-range potential is assumed to be temperature independent. Once the properties of the exchanged particles are temperature dependent, the short-range potential will be temperature dependent, but not significant. In this case, the temperature effect is encoded in the two-body propagator  
\begin{align}
    &G^T(E,m_1,m_2,T)\notag\\
    =&-i\frac{\mu q_0(E,m^T_1,m^T_2)}{8\pi m^T_1m^T_2}\frac{1}{2}\notag\\
    &\times
    \biggl[
    \coth
    \biggl(
    \frac{\beta}{2}
    \biggl(m^T_1+\frac{q_0(E,m^T_1,m^T_2)^2}{2m^T_1}
    \biggr)
    \biggr)
    \notag\\
    &+\coth
    \biggl(
    \frac{\beta}{2}
    \biggl(m^T_2+\frac{q_0(E,m^T_1,m^T_2)^2}{2m^T_2}
    \biggr)
    \biggr)
    \biggr],
    \label{2nrft}
\end{align}
where $m_1^T$ ($m_2^T$) is the mass of the first (second) particle at temperature $T$, i.e., a function of both $m_1$ ($m_2$) and $T$. The details can be found in Appendix~\ref{g2tcal}. Here $\beta=1/T$ is the inverse temperature. In short, the expression at finite temperature can be obtained by replacing the two-body propagator $G(E,m_1,m_2)$ by $G^T(E,m_1,m_2,T)$.

\subsection{The triangle singularity scenario}
\begin{figure}[h]
     \centering
     \includegraphics[width=1\linewidth]{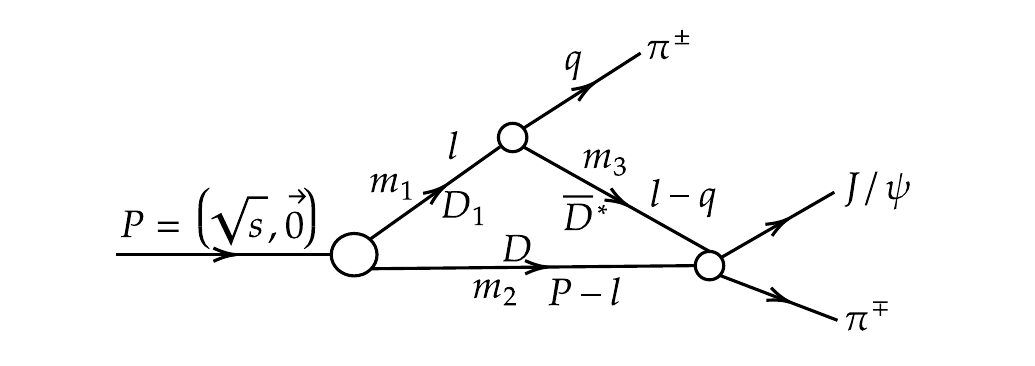}
     \caption{The Feynman diagram corresponding to the production amplitude $\mathcal{M}$ in the triangle singularity scenario. }
     \label{fig_tri}
 \end{figure}

The $Z_c(3900)$ is observed in the $e^+e^-\to J/\psi\pi^+\pi^-$ process at the center-of-mass energy $4.26~\mathrm{GeV}$, which is close to the $D_1\bar{D}$ threshold. On the other hand, $D_1$ primarily decays into D-wave $D^*\pi$. The scattering of the $D^*$ meson and the $\bar{D}$ meson to the $J/\psi\pi$, as illustrated in Fig.~\ref{fig_tri}, induces a peak structure in the $J/\psi\pi$ invariant mass distribution around the $D\bar{D}^*$ threshold, because the triangle singularity condition is satisfied~\cite{Guo:2019twa} at this energy region. The three-point loop function could have singularities, once all the internal particles can be on shell physically as discussed in Ref.~\cite{Coleman:1965xm} and recently reviewed by Ref.~\cite{Guo:2019twa}. The experimental setup where the center-of-mass energy is $4.26$ GeV which is very close to the $D_1 \bar D$ threshold (only slightly below the threshold) and the mass of $D_1$ is larger than the sum of the $D$ and $\pi$ masses (nearly) satisfies the requirement of the above on shell condition (Landau singularity).  
 As a result, the appearance of the $Z_c(3900)$ peak structure in the $J/\psi\pi$ invariant mass distribution is possible~\cite{Wang:2013cya}.
 As all the intermediate particles are charmed mesons, they are dealt with in a nonrelativistic manner.  As a result, the scalar one-loop three-point function reads as~\cite{Guo:2010ak}
 \begin{align}
&I(q,m_1,m_2,m_3)\notag\\
\equiv&i\int \frac{\mathrm{d}^4l}{(2\pi)^4} \frac{1}{(l^2-m_1^2+i\varepsilon)((P-l)^2-m_2^2+i\varepsilon)}
\nonumber \\
&\times\frac{1}{((l-q)^2-m_3^2+i\varepsilon)}\notag\\
\simeq&\frac{-i}{8m_1m_2m_3}\int\frac{\mathrm{d}^4l}{(2\pi)^4}\frac{1}{\left(l^0-m_1-\frac{\vec{l}\,^2}{2m_1}+i\varepsilon\right)}\notag\\
&\times\frac{1}{\left(l^0-M+m_2+\frac{\vec{l}\,^2}{2m_2}-i\varepsilon\right)}\notag\\
&\times\frac{1}{\left(l^0-q^0-m_3-\frac{(\vec{l}-\vec{q})^2}{2m_3}+i\varepsilon\right)}\label{eqstep1}\\
=&\frac{\mu_{12} \mu_{23}}{16 \pi m_1 m_2 m_3} \frac{1}{\sqrt{a}}\biggl[\tan ^{-1}\left(\frac{c^{\prime}-c}{2 \sqrt{a c}}\right)\notag\\
&+\tan ^{-1}\biggl(\frac{2 a+c-c^{\prime}}{2 \sqrt{a\left(c^{\prime}-a\right)}}\biggr)\biggr],
\label{eq:step4}
\end{align}
where
\begin{eqnarray}
    a\equiv\left(\frac{\mu_{23}}{m_3}\right)^2 \vec{q}\,^2,~c\equiv2 \mu_{12} b_{12},~c^{\prime}\equiv2 \mu_{23} b_{23}+\frac{\mu_{23}}{m_3} \vec{q}\,^2,
\end{eqnarray}
and 
\begin{eqnarray}
    b_{12}&\equiv& m_1+m_2-M, \\
    b_{23}&\equiv &m_2+m_3+q^0-M,
\end{eqnarray}
following the conventions of Ref.~\cite{Guo:2010ak}. One can see that the above scalar one-loop function is a logarithmic divergence by replacing the arctan function by the corresponding logarithmic function. When the intermediate particles are unstable, i.e., with sizable widths, the corresponding expression of the above scalar one-loop function can be obtained by the replacement $m_i\to m_i-i\Gamma_i/2$ in the above equations. Here $m_i$ and $\Gamma_i$ are the mass and width of the $i$th particle. In the triangle singularity scenario, the fit function can be obtained by replacing $\bar{\mathcal{M}}$ by the scalar one-loop three-point function. The detailed calculation can be found in Appendix~\ref{I3vacNR}. The temperature effect on the scalar one-loop three-point function is analogous to that of the scalar one-loop two-point function. First, the integral over the energy is replaced by an imaginary variable ($l^0 \to i l^0$), and then the integral is rewritten by a discrete sum over the Matsubara frequency, $z_n = i2 \pi  n/\beta$ ($n\in\textbf{Z}$), to reflect the periodic nature of the Green function at finite temperature. Thus, Eq.~(\ref{eqstep1}) can be computed as
\begin{widetext}
\begin{align}
&I(q,m_1,m_2,m_3)\notag\\
\xrightarrow[]{l^{0}\to il^0}&\frac{1}{8m_1m_2m_3}\int\frac{d^3 l}{(2 \pi)^3}\int \frac{d l^{0}}{2\pi}\frac{1}{(i l^{0}-A_1)(i l^{0}+A_2)(i l^{0}-A_3)}
\nonumber \\
=&\frac{1}{8m^T_1m^T_2m^T_3}\int \frac{d^3 l}{(2 \pi)^3}\frac{1}{\beta}\sum_{z_n=2\pi i n/\beta}\frac{1}{(z_n-A_1)(z_n+A_2)(z_n-A_3)}
\nonumber \\
=&
I^T(q,m^T_1,m^T_2,m^T_3),
\end{align}
\end{widetext}
where the superscript $T$ on the mass is introduced to take into account temperature dependence of the masses, which will be discussed shortly. The sum can be computed by the standard technique of the Sommerfeld-Watson transformation which rewrites it to a contour integral in the complex plane, with the result
\begin{widetext}
\begin{align}
&I^T(q,m^T_1,m^T_2,m^T_3)\notag\\
=&\frac{-1}{8m^T_1m^T_2m^T_3}
\int \frac{d^3 l}{(2 \pi)^3}\biggl(\frac{\coth(A_1\beta/2)}{(A_1+A_2)(A_1-A_3)}
+\frac{\coth(-A_2\beta/2)}{(-A_2-A_1)(-A_2-A_3)}+\frac{\coth(A_3\beta/2)}{(A_3-A_1)(A_3+A_2)}\biggr)\label{eq:85},
\end{align}
\end{widetext}
where
\begin{eqnarray}
    A_1&\equiv& m^T_1+\vec{l}\,^2/(2m^T_1)-i\varepsilon,\\
    A_2&\equiv& m^T_2-M+\vec{l}\,^2/(2m^T_2)-i\varepsilon,\\
    A_3&\equiv& q^0+m^T_3+(\vec{l}-\vec{q})^2/(2m^T_3)-i\varepsilon.
\end{eqnarray}
To obtain the result, we have used the following relations:
\begin{align}
    \frac{1}{\beta}
    \sum_{n=-\infty}^{n=+\infty} f(z_n) =&T \sum_{n=-\infty}^{n=+\infty} f(z_n)
    =-\sum_i \mathrm{Res}(F(z),z=z_i)\notag\\
    =&-\sum_i\frac{1}{2\pi i}\oint_{C_i} d z f(z) \frac{1}{2}
    \coth\biggl(\frac{1}{2}\beta z\biggr),
\end{align}
with 
\begin{eqnarray}
    F(z)\equiv f(z)\frac{1}{2}\coth\biggl(\frac{1}{2}\beta z\biggr),
\end{eqnarray}
and
\begin{eqnarray}
    f(z_n)\equiv \frac{1}{(z_n-A_1)(z_n+A_2)(z_n-A_3)}.
\end{eqnarray}
Notice that $z_n=i2n\pi  /\beta$ is the pole of the $\coth(z\beta/2)$ function, and
$z_i$ is the pole of the $f(z)$ function with $C_i$ the circle around the pole.
It is also noted that we adopt the periodic conditions, i.e., $P^0=i2\pi n_1/\beta$ and $q^0=i2\pi n_2/\beta$ ($n_1,~n_2\in\textbf{Z}$), to the energies of external particles in the argument of the hyperbolic cotangent function in the calculation as done in Ref.~\cite{Gao:2019idb}. 
When the width of the intermediate particle is sizable, one can also obtain the analogous scalar one-loop function with the replacement $m^T_i\to m^T_i-i\Gamma^T_i/2 $. 

\section{Results and Discussions}
\label{sec:results}
\begin{figure}[h]
    \centering
\includegraphics[width=0.8\linewidth]{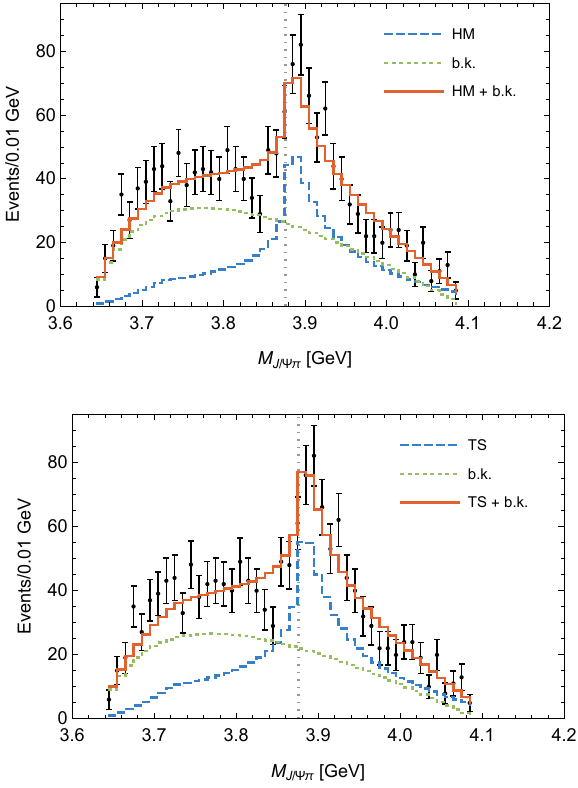}
    \caption{The $Z_c(3900)$ line shape within the hadronic molecular (HM) scenario (upper panel with $\chi^2/\mathrm{d.o.f.}=1.23$) and the triangle singularity (TS) scenario (lower panel with $\chi^2/\mathrm{d.o.f.}=1.33$) in comparison with the 
    invariant mass distribution in Ref.~\cite{BESIII:2013ris}. The blue dashed, green dotted, and red solid curves are the signal, background, and sum of them, respectively. The black points are the experimental data from BESIII~\cite{BESIII:2013ris} at center-of-mass energy $4.26~\mathrm{GeV}$. The vertical line is  the $D\bar{D}^*$ threshold.}
    \label{HMTS_fit}
\end{figure}

Considering that the nature of the $Z_c(3900)$ is still in debate in spite of previous great efforts,
we study whether the various interpretations of the $Z_c(3900)$ exhibit interesting and distinguishable properties at finite temperature. More specifically, we firstly fit the $Z_c(3900)$ line shape in vacuum and investigate its properties at finite temperature. By tuning the parameters in Eq.~\eqref{eq:fit_function} to the experimental data~\cite{BESIII:2013ris}, we obtain reasonable results in both the hadronic molecular scenario and triangle singularity scenario as shown in Fig.~\ref{HMTS_fit}. Here the label of the horizontal axis is $M_{J/\psi \pi}$ because the distributions with two charged pions are the same due to the isospin symmetry approximation. The reduced chi squares are  $\chi^2/\mathrm{d.o.f.}=1.23$ for the hadronic molecular scenario and $\chi^2/\mathrm{d.o.f.}=1.33$ for the triangle singularity scenario, respectively. The fitted parameters are listed in Table~\ref{para_fit}. From Fig.~\ref{HMTS_fit}, we see that both scenarios can describe the data equally well, with almost the same background contribution. This is also consistent with the argument in Ref.~\cite{Pilloni:2016obd}. In this case, we conclude that the current data in vacuum cannot distinguish these two scenarios, which makes us turn to the finite temperature environment.

\begin{figure}[h]
    \centering
    \includegraphics[width=0.8\linewidth]{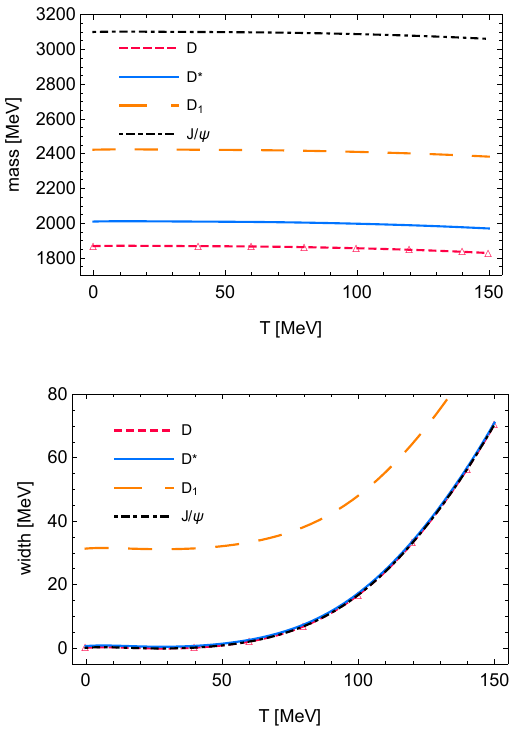}
    \caption{The temperature dependence of masses (upper) and widths (lower) of intermediate particles. The red triangle dots denote the theoretical values of masses and widths 
    for $D$ mesons at different temperatures extracted in Ref.~\cite{Montana:2020lfi}. }
    \label{fig:mass_width}
\end{figure}

The temperature effect on the property of the $Z_c(3900)$ reflects in the hyperbolic cotangent function of the loop function and the intermediate particle masses and widths. The former one can be dealt with numerically. The latter one employs the temperature dependence of the charmed meson in Ref.~\cite{Montana:2020lfi}, which is based on chiral and heavy-quark spin-flavor symmetries in the imaginary time formalism. The mass of the $D$ meson has a sizable decrease as large as $40~\mathrm{MeV}$~\cite{Montana:2020lfi} at temperature $T=150~\mathrm{MeV}$, which agrees with the conclusions in Refs.~\cite{Fuchs:2004fh,Sasaki:2014asa} qualitatively. The dependence of their masses and widths on temperature is presented in Fig.~\ref{fig:mass_width}. The mass (width) decreases (increases) with the increasing temperature. The decreasing mass is because of the chiral restoration at sufficiently high temperature, i.e., the $D$ meson mass should be the same as that of its parity partner with a larger decreasing tendency. The increasing width is because at sufficiently high temperature all the hadrons will be dissociated into quarks and gluons. For other mesons, i.e., $D^*$, $D_1$, and $J/\psi$, we assume their mass and width trajectories are analogous to those of the $D$ meson, but different starting points. With this temperature dependence, we can probe the properties of the $Z_c(3900)$ in the two scenarios.

\begin{figure}[h]
    \centering
    \includegraphics[width=0.8\linewidth]{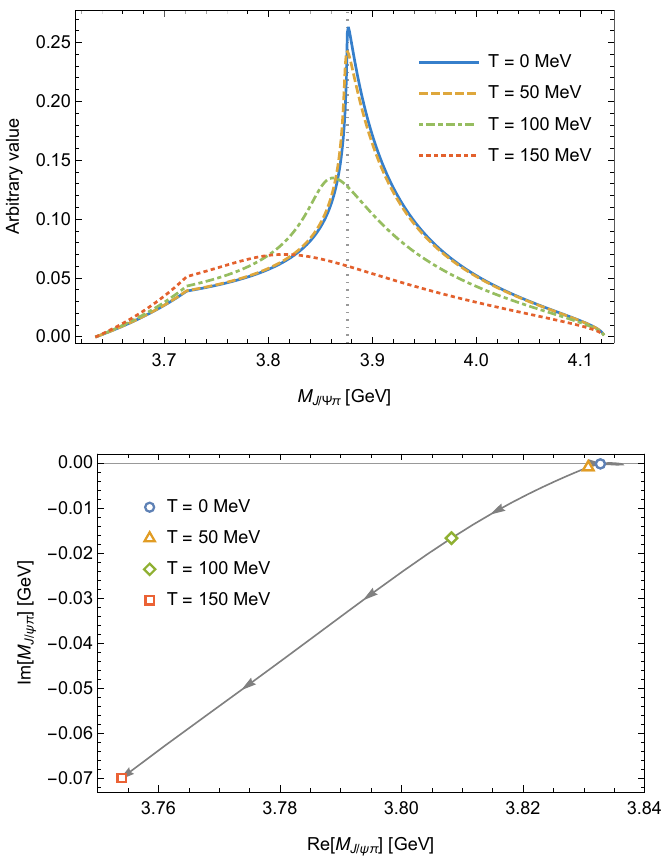}
\captionsetup{justification=raggedright}
    \caption{Upper panel: the line shape of $Z_c(3900)$ in the hadronic molecular scenario at various temperatures. The blue solid, yellow dashed, green dot-dashed and red dotted curves are for the line shape of the $Z_c(3900)$ at temperatures $T=0$ MeV, $50$ MeV, $100$ MeV, and $150$ MeV, respectively. The vertical line indicates the $D\bar{D}^*$ threshold in vacuum. Lower panel: the pole  trajectory when the temperature increases from $T=0~\mathrm{MeV}$ to $T=150~\mathrm{MeV}$. The blue circle, yellow triangle, green rhomboid, and red box indicate the pole positions at temperatures $T=0$ MeV, $50$ MeV, $100$ MeV, and $150$ MeV, respectively.}
    \label{fig:HMtem}
\end{figure}

The line shape of the $Z_c(3900)$ in the hadronic molecular picture at finite temperature is shown in Fig.~\ref{fig:HMtem}. For this scenario, we have assumed that the interaction does not change. We can see that the peak position moves to the lower energy region, when the temperature increases. That is because of the decreasing dependence of the charmed meson mass on temperature (upper panel of Fig.~\ref{fig:mass_width}). The width of the structure becomes wider with the increasing temperature, which stems from the increasing width of the charmed meson (lower panel of Fig.~\ref{fig:mass_width}). The temperature effect from the hyperbolic cotangent function is marginal, because the argument of the hyperbolic cotangent function is so large that it becomes almost unity. This part could contribute significantly to the light system such as pions at low energy~\cite{Gao:2019idb}. The negative thermal mass shift and the increase in the thermal width of $Z_c(3900)$ at high temperature are consistent with those of $X(3872)$ as a molecule $DD^*$ studied in Ref.~\cite{Montana:2022inz}, which calculated the thermal energy of the $X(3872)$ in a pion gas by solving the Lippmann-Schwinger integral equation with charm meson propagators that include
the self-consistent thermal $D^{(*)}$
self energies. 

To get more insight, we make a pole flow on the complex energy plane. We classify Riemann sheets of the two-channel case as
\begin{align}
    R_{++}:\quad \mathrm{Im} ~k_1>0,\quad \mathrm{Im} ~k_2>0,\\
    R_{-+}:\quad \mathrm{Im} ~k_1>0,\quad \mathrm{Im} ~k_2<0,\\
    R_{+-}:\quad \mathrm{Im} ~k_1<0,\quad \mathrm{Im} ~k_2>0,\\
    R_{--}:\quad \mathrm{Im} ~k_1<0,\quad \mathrm{Im} ~k_2<0,
\end{align}
with $k_1$ and $k_2$ the three momenta of the $D\bar{D}^*$ and the $J/\psi\pi$ channels, respectively. We find poles on the $R_{--}$ Riemann sheet as shown in Appendix~\ref{app:pole}. The pole corresponds to the so-called Feshbach resonance, or the quasibound state of $D\bar D^*$ coupled by the lower channel of $J/\psi \pi$. 
With the extracted pole positions, 
we can also extract the corresponding effective couplings to the $J/\psi\pi$ channel ($g_{J/\psi\pi}$) and to the $D\bar{D}^*$ channel ($g_{D\bar{D}^*}$) by 
\begin{align}
    g_{J/\psi\pi}^2(T)=&~\lim_{E\to E_{\mathrm{pole}}}(E-E_{\mathrm{pole}})T_{22}(E,T),\\
 g_{D\bar{D}^*}^2(T)=&~\lim_{E\to E_{\mathrm{pole}}}(E-E_{\mathrm{pole}})T_{11}(E,T).
\end{align}
\begin{figure}[h]
    \centering
    \includegraphics[width=0.8\linewidth]{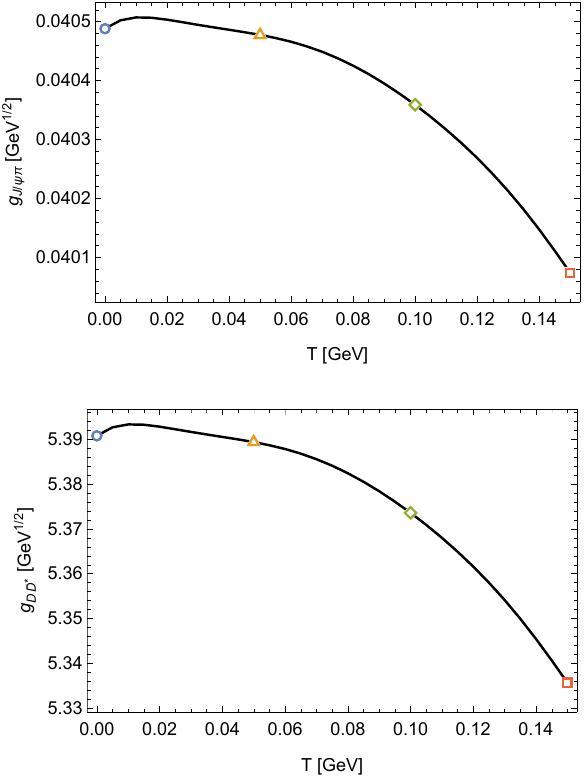}
    \captionsetup{justification=raggedright}
    \caption{The variation of the modulus of effective couplings $g_{J/\psi\pi}$ and $g_{D\bar{D}^*}$ for the poles on the $R_{--}$ Riemann sheet with the increase of temperature $T$. The blue circle, yellow triangle, green rhomboid, and red box indicate the effective couplings at temperatures $T=0$ MeV, $50$ MeV, $100$ MeV, and $150$ MeV, respectively.}
    \label{fig:ectem}
\end{figure}
The extracted couplings are functions of the temperature, and their trajectories are presented in Fig.~\ref{fig:ectem}. It is obvious that the coupling to the $J/\psi\pi$ channel is much smaller than that to the $D\bar{D}^*$ channel, indicating that the $D\bar{D}^*$ is the primary coupling channel. In addition, both these two couplings decrease as the temperature increases, although the amount of the decrease is rather small. It suggests that the hadronic molecule could decouple from all the channels when the temperature increases. 
It also reflects that hadronic molecules would survive in the mass spectrum at sufficiently low temperature, which is the same situation as the $X(3872)$ studied in Refs.~\cite{Cleven:2019cre, Montana:2022inz, Braaten:2023vgs}.

\begin{figure}[h]
    \centering
\includegraphics[width=0.8\linewidth]{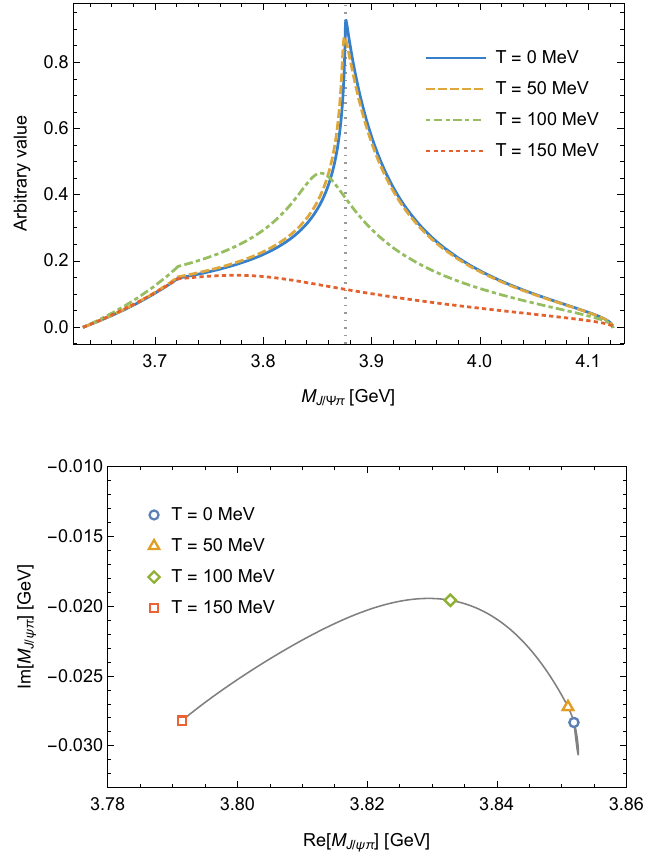}
\captionsetup{justification=raggedright}
    \caption{Upper panel: the line shape of $Z_c$ in the triangle singularity scenario at various temperatures. The blue solid, yellow dashed, green dot-dashed, and red dotted curves are for the line shape of the $Z_c(3900)$ at temperatures $T=0$ MeV, $50$ MeV, $100$ MeV, and $150$ MeV, respectively. The vertical line indicates the $D\bar{D}^*$ threshold in vacuum. Lower panel: the singularity trajectory when the temperature increases from $T=0~\mathrm{MeV}$ to $T=150~\mathrm{MeV}$. The blue circle, yellow triangle, green rhomboid, and red box indicate the triangle singularity positions at temperatures $T=0$ MeV, $50$ MeV, $100$ MeV, and $150$ MeV, respectively. }
    \label{fig:TStem}
\end{figure}

The line shape of the $Z_c(3900)$ in the triangle singularity picture at finite temperature is presented in Fig.~\ref{fig:TStem}. The widths of intermediate particles and their dependence on the temperature (lower panel in Fig.~\ref{fig:mass_width}) are considered in our calculation, which makes the scalar one-loop function finite and the corresponding peak structure becomes wider (upper panel in Fig.~\ref{fig:HMtem}).
The line shape is analogous to that in the hadronic molecular picture, i.e., the peak position moves to the lower energy region when the temperature increases (upper panel of Fig.~\ref{fig:TStem}). That is because the temperature effect mainly comes from the temperature dependence of the masses and widths of the intermediate charmed mesons. The hyperbolic cotangent function for the heavy system is close to 1, and hence the temperature effect is small due to its large mass in comparison with the typical temperature of hadronic degrees of freedom.  The lower panel of Fig.~\ref{fig:TStem} presents the trajectory of the triangle singularity, which is obtained by solving the Landau equation (see Appendix~\ref{Landau}), on the temperature. Although the trajectory behavior is similar to the pole trajectory in the hadronic molecular picture (lower panel of Fig.~\ref{fig:HMtem}), there is still significant deviation due to the different behaviors between the scalar one-loop two-point function and the scalar one-loop three-point function. The imaginary part of the triangle singularity firstly decreases and reaches its minimum value around $T=100~\mathrm{MeV}$. It increases after that temperature. Although the behaviors of the pole trajectory in the hadronic molecular picture and the triangle singularity trajectory are different, the deviation is not sufficient to be distinguished for heavy ion experiments. However, it is worth mentioning that the peaks that originated from these two scenarios are almost smeared when the temperature is larger than $150~\text{MeV}$. It indicates that when the temperature of the environment is high enough, no matter what nature the $Z_c(3900)$ is,
it will be dissociated  and cannot exist anymore. This conclusion is consistent with that in Refs.~\cite{Abreu:2021xpz,Llanes-Estrada:2021ath}.  

\section{Summary}
\label{sec:summary}
We aim at exploring the advantages of studying exotic hadrons in the nonvacuum environment. At the first step, we study the line shape of the $Z_c(3900)$ and its relevant  properties at finite temperature within the hadronic molecular and the triangle singularity pictures. We construct the formulas for these two cases and fit to the  experimental data in vacuum. We find that both of these two scenarios can describe the available experimental line shape equally well, making these two scenarios undistinguishable. Thus, we turn to the nonvacuum case, i.e., the finite temperature case, aiming at the property of the $Z_c(3900)$ at finite temperature and distinguishing the above two scenarios. As we work on the hadron level, the working temperature is limited by the crossover point to quark-gluon degrees of freedom, i.e., about $150~\mathrm{MeV}$. Within this temperature region, the low-energy constants, i.e., the couplings among hadrons and the transition potentials, can be considered temperature independent. The effects of finite temperature are in the change in the masses and decay widths of scattering
hadrons and in the phase space of the loop function (thermal fluctuation of the scattering particles).
The latter results in the hyperbolic cotangent function whose argument is related to the masses of the intermediate mesons and the energies of external particles, which are much larger than the typical temperature of hadron phase, making the temperature effect (thermal fluctuation) insignificant. However, we would expect more significant temperature effect on light systems, especially on low-energy light systems, for instance, the two-pion system which could form a four-quark state $\sigma(500)$. In our case, most of the temperature effect comes from the thermal mass shift and width of the intermediate mesons, which is in agreement with the case in Refs.~\cite{Montana:2022inz, Braaten:2023vgs}. As a result, the line shape of the $Z_c(3900)$ in both the hadronic molecular picture and the triangle singularity picture is almost the same, i.e., the peak structure moves to the lower energy region as the temperature increases. We consider this behavior as a unique character for hadronic interpretations of the $Z_c(3900)$, i.e., the hadronic molecule and the triangle singularity. The width of the structure becomes larger when the temperature increases, indicating that the $Z_c(3900)$ will be dissociated at sufficiently high temperature. This feature can also be seen from the effective couplings in the molecular picture, i.e., a hadronic molecule will decouple from other hadron channels  at high temperature. The marginal difference of these two pictures is from the different behaviors of the two-point and three-point loop functions in terms of the intermediate meson masses. However, owing to the current ability of particle identification in heavy ion experiments, it is difficult to distinguish this marginal difference. The similar behavior of these two scenarios, however,  could be used to extract the temperature of the environment no matter what the origin of the peak structure is.
\begin{widetext}
\begin{center}
\begin{table}[h]
    \caption {The fitted parameters in the hadronic molecular (HM) scenario and the triangle singularity (TS) scenario.}
    \label{para_fit}
    \begin{tabular}{cccccccc}
    \hline
    \hline
    \diagbox{Scenario}{Parameter} & $v$ & $g$ & $a$~[GeV$^b$]   &  $b$  & $c$ & $d~[\mathrm{GeV}^{-1}]$   &  $N$ \\
    \hline
    $\mathrm{HM}$ & $-337.426$&$2.534$ &$7081.093$&$-0.005$ &$-6147.038$&$-221.724$&$2.070\times10^6$  \\
   $\mathrm{TS}$ & & &$4208.752$&$-0.007$ &$-3430.064$&$-185.036$&$8.641\times10^8$  \\
    \hline
    \hline
    \end{tabular}
\end{table}
\end{center}
\end{widetext}
{\bf \color{gray}ACKNOWLEDGMENTS:}~~
We are grateful to Xingyu Guo, Zhiqing Liu, Pengyu Niu, Xinyue Hu, and Quanxing Ye for their helpful discussions. 
This work is partly supported by the National Natural Science Foundation of China with Grants No.~12375073 and NO.~12035007, and Guangdong Provincial funding with Grant No.~2019QN01X172. A.H. is supported in part by the JSPS Grants-in-Aid for Scientific Research with Grants No.~ 21H04478(A) and NO.~24K07050(C). Y.Z. is supported in part by the Guangdong Provincial international exchange program for outstanding young talents of scientific research in 2023.

\newpage
\appendix
\section{Summation of series: Sommerfeld-Watson transformation}\label{AppendixA}
\begin{figure}[h]
    \centering
    \includegraphics[width=1\linewidth]{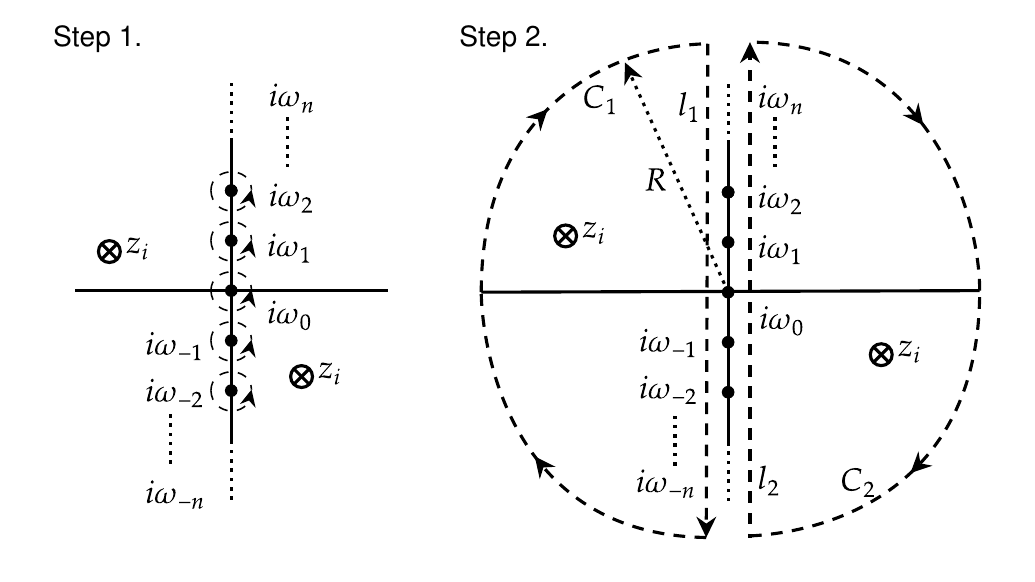}
    \caption{Transformation of the integration contour for Eq.~(\ref{matsubara_sum}).}
    \label{matsubara_sum}
\end{figure}
Assume that $f(z)$ is a function which is analytic at $z=i 2\pi n/\beta$ ($n\in \bm{\mathrm{Z}}$), and tends to zero as fast as, or faster than
\begin{align}
    \frac{1}{|z|^2}~~\text{as}~~|z|\to\infty.
    \label{eq_condition}
\end{align}
 And consider the following sum:
\begin{equation}
    S = \sum^{\infty}_{n=-\infty}f(n).
    \label{matsubara_sum}
\end{equation}
To compute this discrete sum, let us introduce the function $F(z)$ as
\begin{equation}
    F(z)\equiv f(z)\frac{1}{2}\coth\biggl(\frac{z}{2}\beta\biggr)=f(z)\frac{1}{2}\frac{e^{\beta z}+1}{e^{\beta z}-1}.
\end{equation}
This function has simple poles associated with the hyperbolic cotangent function at $z_n=i\omega_n=i2\pi n/\beta$ with residues
\begin{align}
    &\text{Res}(F(z),z=i\omega_n)\notag\\
    =&\lim_{z\to i\omega_n}f(z)\frac{1}{2}\frac{e^{\beta z}+1}{\frac{d}{dz}(e^{\beta z}-1)}\notag\\
    =&\lim_{z\to i\omega_n}f(z)\frac{1}{2}\frac{e^{\beta z}+1}{\beta e^{\beta z}}\notag\\
    =&\frac{1}{\beta}\sum_nf(z_n),
    \label{eq149}
\end{align}
and the poles associated with $f(z)$ at $z=z_i$.

Firstly, we transform the sum of the residues of the poles $z_n = i\omega_n$ to the contour integral over the small circles around these poles as shown on the left of Fig.~\ref{matsubara_sum}. 
Then the sum of the small contours is converted into the sum of the two closed contours of $l_1$ and $C_1$, and of $l_2$ and $C_2$ as shown on the right of Fig.~\ref{matsubara_sum}.
Owing to the condition of Eq.~(\ref{eq_condition}), the contributions of the two semicircles of $C_1$ and of $C_2$ vanish. 
Therefore, we have the relation
\begin{align}
   &\text{Res}(F(z),z=i\omega_n)\notag\\
   =&\frac{1}{2\pi i}\left(\int_{l_1}F(z)dz+\int_{l_2}F(z)dz\right)\notag\\
   =&\frac{1}{2\pi i}\left(\int_{l_1+C_1}F(z)dz+\int_{l_2+C_2}F(z)dz\right)
   \notag\\
   =&-\sum_i\text{Res}(F(z),z=z_i).
    \label{convertion1}
\end{align}
Then finally, one arrives at the following result
\begin{align}
    \frac{1}{\beta}\sum_n f(z_n)=-\sum_i\text{Res}(F(z),z=z_i).
    \label{convertion2}
\end{align}
These manipulations are often called Sommerfeld-Watson transformation, the details of which are, for instance, in Refs.~\cite{Watson:1918ca,Sommerfeld:1949ca}.

\section{The calculation of two-point scalar loop function in nonrelativistic case}\label{g2tcal}
By using the nonrelativistic propagators which are  the inverse of the energy difference, the non-relativistic two-point function in the rest frame of the initial state denoted as $p=(E,\vec{0})$ in vacuum reads as
\begin{align}
    &I_{\text{2NR}}(m_1,m_2,E)\notag\\
    \simeq&\frac{i}{4m_1m_2}\int\frac{d^4l}{(2\pi)^4}\frac{1}{\bigl(l^0-m_1-\vec{l}\,^2/(2m_1)+i\varepsilon\bigr)
    }\notag\\
    &\times\frac{1}{\bigl(E-l^0-m_2-\vec{l}\,^2/(2m_2)+i\varepsilon\bigr)}.
    \label{nr2p}
\end{align}
After performing the contour integral over $l^0$ on the upper half or the lower half plane, we have
\begin{align}
    &I_{2\text{NR}}(m_1,m_2,E)\notag\\
    =&\frac{1}{4m_1m_2}\int\frac{d^3l}{(2\pi)^3}\frac{1}{E-m_1-m_2-\vec{l}\,^2/(2\mu)+i\varepsilon}
    \notag \\
    =&\frac{\mu}{4\pi^2m_1m_2}\int_0^{+\infty}\frac{\vec{l}\,^2dl}{2\mu(E-m_1-m_2)-\vec{l}\,^2+i\varepsilon}
    \notag \\
    \simeq&-\frac{\mu q_0^2}{4\pi^2m_1m_2}\int_0^{+\infty}\frac{dl}{\vec{l}\,^2-q_0^2-i\varepsilon}
    \notag \\
    =&-\frac{\mu q_0^2}{4\pi^2m_1m_2}\int_0^{+\infty}\frac{dl}{(|\vec{l}|-q_0-i\varepsilon)(|\vec{l}|+q_0+i\varepsilon)}
    \notag \\
    =&-\frac{\mu q_0^2}{4\pi^2m_1m_2}\frac{1}{2q_0}\int_0^{+\infty}dl\left(\frac{1}{|\vec{l}|-q_0-i\varepsilon}-\frac{1}{|\vec{l}|+q_0+i\varepsilon}\right)
    \notag \\
    =&-\frac{\mu q_0}{8\pi^2m_1m_2}\biggl(\mathcal{P}\int_0^{+\infty}dl\frac{1}{|\vec{l}|-q_0}+i\pi\delta(|\vec{l}|-q_0)\notag\\
    &-\int_0^{+\infty}dl\frac{1}{|\vec{l}|+q_0+i\varepsilon}\biggr)
    \notag \\
    =&-\frac{\mu q_0}{8\pi^2m_1m_2}\left(\mathcal{P}\int_0^{+\infty}dl\frac{2q_0}{\vec{l}\,^2-q_0^2}+i\pi\right)
    \notag \\
    =&-\frac{\mu q_0}{8\pi^2m_1m_2}(0+i\pi)
    \notag \\
    =&-\frac{i\mu q_0}{8\pi m_1m_2},
    \label{nr2p0}
\end{align}
where $q_0=\sqrt{2\mu(E-m_1-m_2)}$ and $\mu=m_1m_2/(m_1+m_2)$. $\mathcal{P}$ indicates the Cauchy principal value.\\
\indent At finite temperature, after the Sommerfeld-Watson transformation, we use the periodic condition $i\omega_{n}=i2\pi n/\beta~ (n\in\textbf{Z})$ in the argument of $\coth(x)$; then we obtain
\begin{align}
    &I_{2\text{NRFT}}(m_1,m_2,E,T)\notag\\
    =&\frac{1}{4m_1m_2}
    \frac{1}{2}\int\frac{d^3l}{(2\pi)^3}\biggl(\frac{\coth(\beta(m_1+\vec{l}\,^2/(2m_1)
    )/2)}{E-m_1-m_2-\vec{l}\,^2/(2\mu)+i\varepsilon}\notag\\
    &+\frac{\coth(\beta(m_2+\vec{l}\,^2/(2m_2)
    )/2)}{E-m_1-m_2-\vec{l}\,^2/(2\mu)+i\varepsilon}\biggr)\\
    =&\frac{\mu}{4\pi^2m_1m_2}
    \frac{1}{2}\int \vec{l}\,^2dl\biggl(\frac{\coth(\beta(m_1+\vec{l}\,^2/(2m_1)
    )/2)}{2\mu(E-m_1-m_2)-\vec{l}\,^2+i\varepsilon}\notag\\
    &+\frac{\coth(\beta(m_2+\vec{l}\,^2/(2m_2)
    )/2)}{2\mu(E-m_1-m_2)-\vec{l}\,^2+i\varepsilon}\biggr)\\
    \simeq&-\frac{\mu q_0^2}{4\pi^2m_1m_2}
    \frac{1}{2}\int dl\biggl(\frac{\coth(\beta(m_1+q_0^2/(2m_1)-i\varepsilon)/2)}{\vec{l}\,^2-q_0^2-i\varepsilon}\notag\\
    &+\frac{\coth(\beta(m_2+q_0^2/(2m_2)-i\varepsilon)/2)}{\vec{l}\,^2-q_0^2-i\varepsilon}\biggr)\\
    =&-\frac{i\mu q_0}{8\pi m_1m_2}
    \frac{1}{2}\biggl[
    \coth\biggl(\frac{\beta}{2}\biggl(m_1+\frac{q_0^2}{2m_1}
    \biggr)
    \biggr)
    \notag\\
    &+\coth\biggl(\frac{\beta}{2}\biggl(m_2+\frac{q_0^2}{2m_2}
    \biggr)
    \biggr)
    \biggr].
    \label{nr2pft}
\end{align}
\section{The calculation of the nonrelativistic three-point scalar loop function in vacuum}\label{I3vacNR}
We calculate the loop function in vacuum as
\begin{widetext}
\begin{eqnarray}
&~&I(q,m_1,m_2,m_3)\notag\\
&\equiv&i\int \frac{\mathrm{d}^4l}{(2\pi)^4} \frac{1}{(l^2-m_1^2+i\varepsilon)((P-l)^2-m_2^2+i\varepsilon)((l-q)^2-m_3^2+i\varepsilon)} \notag \\
&\simeq&\frac{i}{8 m_1 m_2 m_3}\int \frac{\mathrm{d}^4l}{(2\pi)^4} \frac{1}{(l^0-m_1-\frac{\vec{l}\,^2}{2m_1}+i\varepsilon)(P^0-l^0-m_2-\frac{\vec{l}\,^2}{2m_2}+i\varepsilon)(l^0-q^0-m_3-\frac{\left(\vec{l}-\vec{q}\right)^2}{2m_3}+i\varepsilon)} \notag \\
&\xrightarrow[]{l^{0}-m_1\to l^0}&\frac{-i}{8 m_1 m_2 m_3} \int \frac{d^4 l}{(2 \pi)^4} \frac{1}{\left(l^0-\frac{\vec{l}\,^2}{2m_1}+i \varepsilon\right)\left(l^0+b_{12}+\frac{\vec{l}\,^2}{2m_2}-i \varepsilon\right)\left[l^0+b_{12}-b_{23}-\frac{\left(\vec{l}-\vec{q}\right)^2}{2m_3}+i \varepsilon\right]}\label{eq:step1} \\
& =&\frac{\mu_{12} \mu_{23}}{2 m_1 m_2 m_3} \int \frac{d^{d-1} l}{(2 \pi)^{d-1}} \frac{1}{\left(\vec{l}\,^2+c-i \varepsilon\right)\left(\vec{l}\,^2-\frac{2 \mu_{23}}{m_3} \vec{l} \cdot \vec{q}+c^{\prime}-i \varepsilon\right)} \label{eq:step2}\\
& =&\frac{\mu_{12} \mu_{23}}{2 m_1 m_2 m_3} \int \frac{d^{d-1} l}{(2 \pi)^{d-1}} \frac{1}{\left(\vec{l}\,^2+c-i \varepsilon\right)\left(\vec{l}\,^2-2\sqrt{a}\cos\theta|\vec{l}| +c^{\prime}-i \varepsilon\right)} \notag \\
& =&\frac{\mu_{12} \mu_{23}}{2 m_1 m_2 m_3} \int_0^1 d x \int \frac{d^{d-1} l}{(2 \pi)^{d-1}} \frac{1}{\left[\vec{l}\,^2-a x^2+\left(c^{\prime}-c\right) x+c-i \varepsilon\right]^2} \notag\\
& =&\frac{\mu_{12} \mu_{23}}{2 m_1 m_2 m_3} \int_0^1 d x \frac{i}{(4\pi)^{3/2}}\frac{\Gamma(1/2)}{\Gamma(2)}\Delta^{-1/2} \label{eq:step7} \\
& =&\frac{\mu_{12} \mu_{23}}{16 \pi m_1 m_2 m_3} \frac{1}{\sqrt{a}} \log\left( -(c^\prime-c)+2a x+2\sqrt{a}\sqrt{-c-(c^\prime-c)x+a x^2}\right)_{x=0}^{x=1} \notag\\
& =&\frac{\mu_{12} \mu_{23}}{16 \pi m_1 m_2 m_3} \frac{1}{\sqrt{a}}\left[\tan ^{-1}\left(\frac{c^{\prime}-c}{2 \sqrt{a c}}\right)+\tan ^{-1}\left(\frac{2 a+c-c^{\prime}}{2 \sqrt{a\left(c^{\prime}-a\right)}}\right)\right],\notag
\end{eqnarray}
\end{widetext}
where
\begin{eqnarray}
    a\equiv\left(\frac{\mu_{23}}{m_3}\right)^2 \vec{q}\,^2,~c\equiv2 \mu_{12} b_{12},~ c^{\prime}\equiv2 \mu_{23} b_{23}+\frac{\mu_{23}}{m_3} \vec{q}\,^2,
\end{eqnarray}
and 
\begin{eqnarray}
    b_{12}&\equiv& m_1+m_2-M,\\
    b_{23}&\equiv &m_2+m_3+q^0-M,
\end{eqnarray}
following the conventions of Ref.~\cite{Guo:2010ak}. Note that in Eq.~\eqref{eq:step1}, we perform a replacement $l^0-m_1\to l^0$, as the integration over $l^0$ is convergent. In Eq.~\eqref{eq:step2}, the integration over the upper-half plane is performed. The result of the integration over the lower-half plane is the same. After finishing the integral over $l$, the dimension $d$ is set $d = 4$ to obtain the result~\eqref{eq:step7}.
\section{The application of the Laudau equation to the triangle singularity}\label{Landau}
\begin{figure}[h]
    \centering
    \includegraphics[width=1\linewidth]{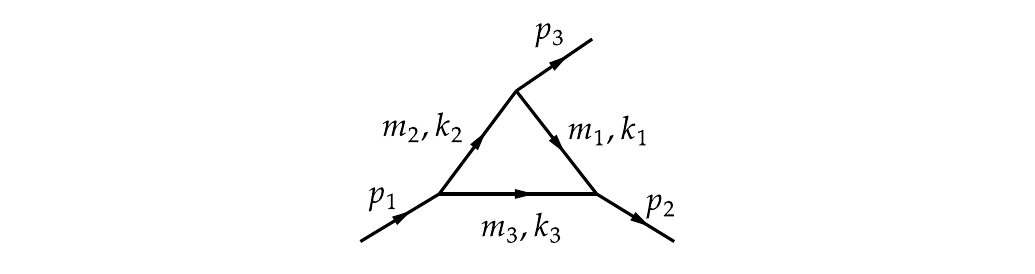}
    \caption{The momentum convention of the three-point loop.}
    \label{landau_eq}
\end{figure}

The Landau equation of the three-point loop can be obtained from the following equations:
  \begin{align}  &\alpha_1k_1^{\mu}+\alpha_2k_2^{\mu}+\alpha_3k_3^{\mu}=0,\label{eq:E1}\\
    &k_1^2-m_1^2=k_2^2-m_2^2=k_3^2-m_3^2=0, \label{eq:E2} 
  \end{align} 
where $\alpha_i$ is the Feynman parameter, and $k_i$ and $m_i$ are the four momentum and mass of each intermediate particle as shown in Fig.~\ref{landau_eq}.
\\
\indent Contracting Eq.~\eqref{eq:E1} with $k_{1\mu}$, $k_{2\mu}$ and $k_{3\mu}$, respectively, and using Eq.~\eqref{eq:E2} one can obtain
\begin{align}
    &\beta_1+\beta_2y_{12}+\beta_3y_{13}=0,\\
    &\beta_1y_{12}+\beta_2+\beta_3y_{23}=0,\\
    &\beta_1y_{13}+\beta_2y_{23}+\beta_3=0,
\end{align}
with $\beta_i\equiv\alpha_im_i$, and 
\begin{equation}
    y_{ij}\equiv\frac{k_i\cdot k_j}{m_im_j}=\frac{m_i^2+m_j^2-p_k^2}{2m_im_j}~(i\neq j\neq k),
    \label{yij}
\end{equation}
where $p_k$ is the external four momentum. The sufficient and necessary condition for a solution of the above equations is 
\begin{equation}
\left | \begin{matrix}
1 &y_{12}   & y_{13} \\
y_{12} & 1 &y_{23} \\
y_{13} &y_{23} & 1  \\
\end{matrix} \right |=1+2y_{12}y_{23}y_{13}-y_{12}^2-y_{23}^2-y_{13}^2=0, 
\end{equation}
which determines the location of the leading triangle singularities. For a detailed review of the triangle singularity, we recommend Ref.~\cite{Guo:2019twa}.
\section{Pole positions and the corresponding effective couplings at various temperatures}
\label{app:pole}
Table~\ref{hm_pole} shows the pole positions at various temperatures in the hadronic molecular scenario. Please note that the negative imaginary parts of the values in the former ten rows are so small that they are omitted in this table. Table~\ref{hm_coupling} shows the modulus of the corresponding effective couplings $(g_{J/\psi\pi},g_{D\bar{D}^*})$ at various temperatures in the hadronic molecular scenario.
\begin{center}
\begin{table}[h]
    \caption {The pole positions in GeV unit at various temperatures in the hadronic molecular scenario. Poles appear on the Riemann sheets $R_{+-}$ and $R_{--}$.}
    \label{hm_pole}
    \begin{tabular}{c c c}
    \hline
    \hline
    \diagbox{$T~\mathrm{[MeV]}$}{\text{R.S.}}~~~~~~~ &  ~~~~~~~$R_{+-}$~~~~~~~    &  ~~~~~~~$R_{--}$~~~~~~~ \\[1pt]
    \hline
    $0$   &$3.833$&    $3.833$\\
    $5$   &$3.835$&    $3.835$\\
    $10$   &$3.836$&    $3.836$\\
    $15$   &$3.836$&    $3.836$\\
    $20$   &$3.836$&    $3.836$\\
    $25$   &$3.835$&    $3.835$\\
    $30$   &$3.834$&    $3.834$\\
    $35$   &$3.833$&    $3.833$\\
    $40$  &$3.832$ &  $3.832$ \\
    $45$   &$3.832$ &   $3.832$ \\
    $50$  &$3.831-0.001i$ &   $3.831-0.001i$ \\
   $55$  &$3.830-0.002i$ &   $3.830-0.002i$\\
   $60$  &$3.829-0.002 i$ &  $3.829-0.002 i$\\
   $65$  &$3.827-0.003i$&   $3.827-0.003i$ \\
    $70$   &$3.825-0.004i$&   $3.825-0.004i$ \\
   $75$  &$3.823 -0.005i$ &  $3.823 -0.005i$\\
   $80$  &$3.821 -0.007 i$ &   $3.821 -0.007 i$\\
   $85$   &$3.818-0.009 i$&    $3.818-0.009 i$\\
    $90$   &$3.815-0.011i$&    $3.815-0.011i$\\
    $95$  &$3.812-0.014i$ &    $3.812-0.014i$ \\
    $100$  &$3.808-0.017 i$ &    $3.808-0.017 i$\\
    $105$  &$3.804-0.020 i$ &   $3.804-0.020 i$\\
    $110$  &$3.800-0.024 i$ &   $3.800-0.024 i$\\
    $115$  &$3.796 -0.028  i$ &   $3.796 -0.028  i$\\
    $120$  &$3.791-0.033 i$ &   $3.791-0.033 i$\\
    $125$  &$3.786-0.038i$ &  $3.786-0.038i$ \\
    $130$  &$3.780-0.044 i$  &  $3.780-0.044 i$ \\
    $135$  &$3.774-0.050 i$ &    $3.774-0.050 i$\\
    $140$  &$3.768-0.056 i$ &   $3.768-0.056 i$ \\
    $145$  &$3.761-0.063i$ &   $3.761-0.063i$ \\
    $150$  &$3.754-0.070i$ &   $3.754-0.070i$ \\
    \hline
    \hline
    \end{tabular}
\end{table}
\begin{table}[t]
    \caption {The modulus of effective couplings $(g_{J/\psi\pi},g_{D\bar{D}^*})$ in GeV$^{1/2}$ unit at various temperatures in the hadronic molecular scenario.}
    \label{hm_coupling}
    \begin{tabular}{c c c}
    \hline
    \hline
    \diagbox{$T~\mathrm{[MeV]}$}{\text{R.S.}}~~~~~~~ &  ~~~~~~~$R_{+-}$~~~~~~~    &  ~~~~~~~$R_{--}$~~~~~~~ \\
    \hline
    $0$   &$(0.040,5.391)$&   $(0.040,5.391)$ \\
    $5$   &$(0.041,5.393)$&   $(0.041,5.393)$ \\
    $10$   &$(0.041,5.393)$&   $(0.041,5.393)$ \\
    $15$   &$(0.041,5.393)$&   $(0.041,5.393)$ \\
    $20$  &$(0.041,5.393)$ &  $(0.041,5.393)$ \\
    $25$   &$(0.040,5.392)$&   $(0.040,5.392)$ \\
    $30$   &$(0.040,5.392)$&   $(0.040,5.392)$ \\
    $35$   &$(0.040,5.391)$&   $(0.040,5.391)$ \\
   $40$  &$(0.040,5.391)$ &  $(0.040,5.391)$ \\
    $45$  &$(0.040,5.390)$ &  $(0.040,5.390)$ \\
    $50$  &$(0.040,5.389)$ &   $(0.040,5.389)$ \\
    $55$  &$(0.040,5.389)$ &   $(0.040,5.389)$\\
    $60$  &$(0.040,5.388)$ &   $(0.040,5.388)$\\
    $65$  &$(0.040,5.387)$  &  $(0.040,5.387)$ \\
    $70$  &$(0.040,5.386)$ &   $(0.040,5.386)$\\
    $75$ &$(0.040,5.384)$ &   $(0.040,5.384)$ \\
    $80$   &$(0.040,5.382)$&   $(0.040,5.382)$ \\
   $85$  &$(0.040,5.381)$ &  $(0.040,5.381)$ \\
    $90$  &$(0.040,5.378)$ &   $(0.040,5.378)$ \\
    $95$  &$(0.040,5.376)$ &   $(0.040,5.376)$\\
    $100$  &$(0.040,5.374)$ &   $(0.040,5.374)$\\
    $105$  &$(0.040,5.371)$  &  $(0.040,5.371)$ \\
    $110$  &$(0.040,5.368)$ &    $(0.040,5.368)$\\
    $115$  &$(0.040,5.365)$ &   $(0.040,5.365)$ \\
    $120$  &$(0.040,5.362)$ &  $(0.040,5.362)$ \\
    $125$  &$(0.040,5.358)$ &   $(0.040,5.358)$ \\
    $130$  &$(0.040,5.354)$ &   $(0.040,5.354)$\\
    $135$  &$(0.040,5.350)$ &   $(0.040,5.350)$\\
    $140$  &$(0.040,5.345)$  &  $(0.040,5.345)$ \\
    $145$  &$(0.040,5.341)$ &   $(0.040,5.341)$\\
    $150$  &$(0.040,5.336)$ &   $(0.040,5.336)$\\ 
    \hline
    \hline
    \end{tabular}
\end{table}
\end{center}

\end{document}